\documentclass[aps,prl,twocolumn,showpacs,a4paper]{revtex4}
\usepackage{graphicx}

\begin{document}





\title{Continuum Theory of Tkachenko Modes in Rotating Bose--Einstein Condensate}
\author{E.B. Sonin}

\affiliation{Racah Institute of Physics, Hebrew University of
Jerusalem, Jerusalem 91904, Israel   }

\date{\today}

\begin{abstract}

The present paper suggests the continuum theory of Tkachenko modes in a rotating 2D  Bose--Einstein
condensate taking into account density inhomogeneity  and compressibility of the condensate. The problem
requires solution of coupled hydrodynamic equations for vortex and liquid motion with proper boundary
conditions, which were derived for the condensate described by the  Thomas--Fermi approximation. Compressibility
becomes essential at rapid rotation with angular  velocity close to the trap frequency. The theory is in a
reasonable agreement with experimental observation of Tkachenko modes.
\end{abstract}



%



\maketitle

Tkachenko mode \cite{Tkach} is a transverse sound wave in a vortex lattice, which exists in the ground state of a
rotating superfluid. It has attracted a lot of attention in vortex
dynamics of laboratory and astrophysical superfluids. Observation of this extremely soft mode  (of order
or less than a few  Hz) in laboratory superfluids was very difficult because even slight pinning of vortex ends
shades the contribution of vortex shear rigidity transforming the Tkachenko wave into a classical inertial wave 
\cite{RMP}. However, discovery of Bose--Einstein condensate (BEC) and possibility to rapidly rotate it made
observation of Tkachenko modes more feasible, which resulted in clear experimental detection of them \cite{exp}.
As well as the Kelvin mode, also possible in the BEC \cite{Fett}, the Tkachenko mode is a direct manifestation
of the quantum vorticity since it depends on the circulation quantum.

In a number of aspects external conditions for Tkachenko waves in BEC are essentially different from those
in ``old'' superfluids. First, in a rapidly rotating BEC one cannot consider a superfluid to be
incompressible. Effect of finite compressibility was investigated theoretically in the past \cite{RMP} since the
effect is crucial for the long-wavelength limit of the Tkachenko mode. But at that time it was considered as a
theoretical curiosity since in order to reveal it one needed containers about a few hundred meters diameter. In the
BEC case the situation is essentially different because in contrast to a strongly interacting bose liquid as He
II, BEC is a weakly interacting bose gas with very low sound speed and very high compressibility. Importance of
high liquid compressibility for Tkachenko waves in BEC was pointed out by Baym \cite{Baym}, who rederived the
spectrum of Tkachenko waves in a compressible liquid \cite{Pit} known from Ref. \onlinecite{RMP} and compared it
with the experiment
\cite{exp}. Another important feature of rotating BEC, also connected with its high compressibility, is that the
liquid density is essentially inhomogeneous. This feature was taken into account in the theory by Anglin and
Crescimanno \cite{Ang} within the frame of the continuum theory, which replaces a discrete vortex
lattice by a continuous medium, like the elasticity theory for atomic crystals. But they neglected liquid compressibility,
while a proper comparison with the experiment requires a theory taking into account the both features, compressibility
and inhomogeneity. Recently the Tkachenko mode in rotating BEC was investigated numerically with solving the
equations of Gross--Pitaevskii theory (mean-field theory) \cite{Jap,Bak}. The numerical results well agreed with
the experiment. But numerical calculations, whatever powerful and useful they can be, do not exclude necessity to
develop the analytical approach,  as far as it possible, since the latter gives analytical and scaling
properties and therefore provides a deeper insight into physics of the phenomenon. In the experiment there are
good conditions for application of a continuum theory since usually the relevant length scale (the condensate
size) essentially exceeded the intervortex distance. The goal of the present paper is to suggest the continuum
theory of Tkachenko modes in a rapidly rotating BEC taking into account liquid compressibility and
inhomogeneity. The main challenge for the theory was to formulate proper boundary conditions for oscillating BEC.

Let us remind the equations of motion of a homogeneous compressible 
superfluid at $T=0$ (no normal component) in the rotating coordinate frame \cite{RMP}:
\begin{equation}
{\partial \rho'\over \partial t}+\rho_0 \vec \nabla \cdot \vec v=0~,
\end{equation}
\begin{equation}
\rho_0\left({\partial \vec v\over \partial t}+ 2\vec \Omega \times \vec v_L\right)= -c_s^2\vec
\nabla \rho'~,
\end{equation}
\begin{equation}
\rho_0\vec \kappa \times (\vec v_L-\vec v)={\kappa c_T^2 \over 2\Omega}[2 \vec \nabla (\vec \nabla \cdot
\vec u)- \Delta \vec u]~.
\end{equation}
Here $\rho'(\vec r)$ is the oscillating component of the liquid mass density $\rho(\vec r)=\rho_0 + \rho'(\vec
r)$ around  the equilibrium homogeneous density $\rho_0$, $\vec u$ is the vortex displacement, $\vec
v_L=d\vec u/dt$ is the vortex velocity, $\vec v$ is the liquid velocity averaged over the vortex-lattice cell,
$\vec \Omega$ is the angular velocity, $c_s$ is the sound velocity, $\kappa=h/m$ is the circulation quantum, and
$c_T=\sqrt{\kappa \Omega/8\pi}$ is the Tkachenko-wave velocity. Assuming that $c_T\ll c_s$ and $c_T k \ll \Omega$
one receives for the spectrum of plane waves $\propto \exp(i\vec k \cdot \vec r-i\omega t)$ 
the dispersion relation:
\begin{equation}
\omega^4- \omega^2( 4\Omega^2+c_s^2k^2) +c_s^2c_T^2k^4=0~,
       \end{equation}
which yields the gapped sound mode
\begin{equation}
\omega^2= 4\Omega^2+c_s^2k^2
     \label{SM}  \end{equation}
and the soft quantum mode (see  Ref. \onlinecite{Vol} and Eq. (4.73) in Ref. \onlinecite{RMP})
\begin{equation}
\omega^2={c_T^2c_s^2k^4 \over 4\Omega^2+c_s^2k^2}~.
      \label{TM} \end{equation}
So for the soft (Tkachenko) mode compressibility is essential in the long-wavelength limit $k\ll
\Omega/c_s$. It transforms the Tkachenko wave with the sound spectrum $\omega=c_Tk$ to a softer mode $\omega \propto
k^2$.  In the Tkachenko mode motion of both the vortex lattice and the liquid  has an elliptic polarization but
longitudinal components  parallel to the wave vector
$\vec k$ are very small. The transverse components (normal to $\vec k$) of the vortex lattice and the liquid are 
close one to another: $v_{Lt} \approx v_t$.

Generalization of these equations onto an inhomogeneous liquid with equilibrium density $\rho_0(\vec r)$, which
varies in the plane normal to the rotation axis. is straightforward. The goal of this paper is to consider only
axisymmetric Tkachenko  eigenmodes in a axisymmetric rotating BEC. Thus we write the equations of motion in the
polar system of coordinates for the monochromatic mode $\propto e^{-i\omega t}$:
\begin{equation}
2 \Omega i\omega v_r=-\omega ^2 v_t-{c_T^2\over \rho_0} {1\over r^2}{\partial \over \partial
r}\left[ \rho_0r^3{\partial \over \partial r}\left(v_t\over r\right)\right]~, 
        \label{eq1}\end{equation}
\begin{equation}
 2 \Omega i\omega v_t= {c_s^2\over \rho_0}{\partial \over \partial r}\left[ {1\over r}{\partial
(\rho_0 rv_r)\over \partial r}\right] ~.
    \label{eq2}\end{equation}
We excluded all variables except for the
tangential (azimuthal in polar coordinates) component of the velocity $v_t\approx v_{Lt}$ and the radial
component of the liquid velocity $v_r$. The latter, though being much smaller than $v_t$, is crucial for the
compressibility effect. For a weakly interacting bose gas $c_s^2$ is proportional  to the density $\rho_0$. Therefore
the ratio $c_s^2/\rho_0$ can be replaced by its value $c_s^2(0)/\rho_0(0)$ in the BEC center $r=0$.

Let us remind the results for Tkachenko eigenmodes in an incompressible ($c_s\rightarrow \infty$) homogeneous
($\rho_0=const$) liquid in a cylindric container. Then the radial  liquid velocity $v_r$ vanishes, and Eq.
(\ref{eq1}) is the Bessel equation with a solution
$v_t \propto J_1(kr)$. The eigenvalues of the wave number $k$ are determined from the boundary condition at
$r=R$, where
$R$ is the radius of the vortex-lattice sample, which normally is close to the container radius. In
the ideal axisymmetric case without any interaction of vortices with lateral walls the radial flux of the
transverse momentum of the vortex lattice, which is  given by the corresponding component of the lattice stress
tensor, must vanish. This gives the boundary condition
\begin{equation}
{du(R)\over dr} -{u(R)\over R}=-{1\over i\omega }\left[{dv_t(R)\over dr} -{v_t(R)\over R}\right]=0~.
     \label{BC}  \end{equation}
The flux of the 
 angular moment into the liquid is proportional to the same stress-tensor component and therefore vanishes also.
Equation (\ref{BC}) leads to the condition $J_2(kR)=0$ imposed on the wave numbers $k$. Traditionally in the
papers on BEC they scale the Tkachenko-mode frequencies by the frequency $\Omega b/R$, where
$b=\sqrt{\kappa/\sqrt{3} \Omega}$ is the intervortex distance \cite{exp,Ang,Bak}:  
\begin{equation}
\omega=  \tilde \omega {\Omega b\over R}= \tilde \omega \sqrt{8\pi\over \sqrt{3}}{c_T\over R}~.
            \end{equation}
Then the first two roots $kR=5.14$ and 8.42 of the equation $J_2(kR)=0$ give the first two reduced eigenfrequencies of
the Tkachenko mode
$\tilde
\omega =\sqrt{\sqrt{3}/8\pi }kR= 0.263 kR$ with the ratio
8.42/5.14=1.64.
Another interesting case is the limit of strong interaction of vortices with rough lateral walls, when
$v_t(R)=0$. Then the eigenfrequencies are given by two first roots of the equation
$J_1(kR)=0$ $kR=3.83$ and 7.02  with the ratio of  the two lowest eigenfrequencies  7.02/3.83=1.83. 

Now let us consider an inhomogeneous (but still incompressible) liquid. The analysis will be applied to the rotating
2D BEC in pancake geometry with the density profile $\rho_0 =\rho(0)(1-r^2/R^2)$ determined from the
Thomas--Fermi approximation \cite{Dalf}. Here
$R=\sqrt{2}c_s(0)/\sqrt{\omega_\perp^2-\Omega^2}$ is the BEC radius (Thomas--Fermi radius),  $c_s(0)$ is the
sound velocity at the symmetry axis
$r=0$, and $\omega_\perp$ is the trap frequency, which characterizes the curvature of the trapping parabolic
potential \cite{exp}. Since the momentum flux is proportional to
$\rho_0$ and the latter vanishes at $r=R$, it looks that the flux through the boundary vanishes independently on
whether the boundary condition Eq. (\ref{BC}) is satisfied or not and the latter is unnecessary. But this is not the
case as one can see from solution of the equation of motion close to $r=R$ by expansion in small $(R-r)/R$. With
the relative error of the order of $(R-r)^2/R^2$ the general solution of Eq. (\ref{eq1}) is $v_t \approx
r[C_1 +C_2 \ln (R-r)]$, where $C_1$ and $C_2$ are arbitrary constants. The component $\propto C_2$ gives a divergent
contribution to the displacements and a finite contribution to the stress tensor despite the factor $\rho_0
\propto R-r$. This would violate the  conservation law  for the angular moment and this component should be
absent. This requirement is satisfied if the boundary condition Eq. (\ref{BC}) takes place. Therefore the
condition Eq. (\ref{BC}) is relevant also for an inhomogeneous  liquid with
$\rho_0(r) \rightarrow 0$ at the border. The equation of motion for an inhomogeneous  liquid has no
analytic solution in the whole interval of $r$ but it is not difficult using Mathematica to find the eigenmodes. First
two of them are $\tilde
\omega=1.428$ and 2.327  with the ratio of  the two lowest eigenfrequencies  2.327/1.428=1.63. This agrees with
the results of Ref. \onlinecite{Ang}.

According to Eq. (\ref{TM}) the compressibility effect becomes important if $k$  is of the order or
less than $\Omega/c_s$. Since the eigenvalues of $k$ are of the order of $1/R$ this yields the
condition for a strong compressibility effect:  $\Omega R/c_s(0) \sim \Omega/\sqrt{\omega_\perp^2-\Omega^2} \gg
1$. Thus at rapid rotation of the BEC with angular velocity $\Omega$ close to the trap velocity $\omega_\perp$.
liquid compressibility should be taken into account.
If we want to make the quantitative  analysis for an incompressible liquid, we should solve the system of
two coupled  second-order differential equations (\ref{eq1}) and (\ref{eq2}) with proper boundary conditions.
One  of them remains to be Eq. (\ref{BC}), but we need also the boundary condition imposed on the radial liquid
velocity $v_r$. We use the arguments similar to those used for derivation of Eq. (\ref{BC}). The total mass
balance requires that the radial mass current
$\rho_0(r) v_r(r)$ at the BEC border  $r=R$ vanishes. Indeed, even though the BEC border is mobile because of the
oscillation, the total mass, which can be transferred through the equilibrium border $r=R$,  is a second-order
quantity with respect to an oscillation amplitude. The radial mass
current vanishes for any finite radial velocity $v_r(R)$ since
$\rho(r) \rightarrow 0$ at $r\rightarrow R$. But some condition is required to provide $v_r(R)$ to be finite. In order
to derive this condition we solve Eq. (\ref{eq2}) at $r \approx R$ by series expansion [neglecting terms $\sim
(R-r)^2$]:
\begin{equation}
 v_r(r)= { \Omega i\omega v_t(R)R\over c_s(0)^2}{R-r\over 2}+C_1 \left(1+{R-r
\over R}\right)+{C_2\over R-r}~.
   \end{equation}
The divergent component $\propto C_2$  gives a finite mass flow at the border and should be ruled out. Taking a
derivative from this expression and excluding the constant $C_1$ we receive the boundary condition imposed on $v_r$:
\begin{equation}
{dv_r(R)\over d r}+{v_r(R)\over R}= -{i\omega\Omega R\over 2 c_s(0)^2}v_t(R)~.
  \label{BC-2} \end{equation}

The left-hand side of the boundary condition (\ref{BC-2}) is the divergence $\vec \nabla \cdot \vec v$ of the liquid
velocity in polar  coordinates. The divergence remains constant at the  boundary of the BEC  sample, but
the density $\rho_0(R)$ vanishes there. As a result, the mass continuity equation yields that 
\begin{equation}
 {\partial \rho \over \partial t} = -\vec \nabla \cdot (\rho_0 \vec v) = - \vec v \cdot  \vec \nabla \rho_0~.
   \end{equation} 
Thus at the sample boundary the whole variation of the density originates from the oscillation of the boundary with
the velocity $v_r(R)$. The same boundary condition is valid for other compressional modes of the BEC
\cite{Str,Dal}.

 For better understanding of parametric
dependences of the eigenfrequencies it is useful to introduce the dimensionless radius $z=r/R$ and to scale the radial
velocity, introducing the velocity $u$: $v_r=iuc_T/c_s(0)$. Then the system of equations is:
\begin{eqnarray}
sq(1-z^2) u(z) = q^2 (1-z^2) v_t(z)\nonumber \\
  +{1\over z^2}{d \over dz}\left\{ (1-z^2)z^3{d \over
dz}\left[v_t(z)\over z\right]\right\} ~,
      \label{eqD1}  \end{eqnarray}
\begin{equation}
 s q v_t(z)={d \over dz}\left\{ {1\over z}{d
[(1-z^2) zu(z)]\over dz}\right\} ~,
   \label{eqD2}\end{equation}
where $q$ determines the reduced frequency, $\tilde \omega =0.263 q$\,,  and the parameter 
\begin{equation}
 s ={2\Omega R \over c_s(0) }={2\sqrt{2}\Omega  \over
\sqrt{\omega_\perp^2-\Omega^2}}
   \end{equation} 
characterizes the effect of compressibility. In the dimensionless form the boundary condition (\ref{BC-2}) is
$du(1)/dz+u(1)=-sqv_t(1)/4$.

Let us discuss now restrictions on validness of the presented analysis. In order to observe the strong
compressibility effect $s\gg 1$ one should rotate the BEC rapidly, at $\Omega \sim \omega_\perp$.  But if $\Omega $
approaches  to $\omega_\perp$ close enough at fixed number of atoms the condition of small ratio $c_T/c_s$ is
violated. Since the vortex core size is $\xi \sim \kappa/ c_s$, the large ratio $c_T/c_s \sim \xi/b$ means that
the vortex cores start to overlap. This signals that we approach to  the critical angular velocity, similar to
the upper critical magnetic field $H_{c2}$, when rotation should suppress the superfluid order parameter.  When
$\xi \sim b$ the properties of the BEC become essentially different from those assumed in the present work (see
Ref. \onlinecite{Cod} and references therein). However with a  number of atoms large enough one can reach very
high values of $s$ without violation of the condition $\xi/b \ll 1$ \cite{comm}. 
Another assumption of our analysis was solid body
rotation of the BEC with the constant vortex density given by the Feynman formula
$n_v=2\Omega/\kappa$.  In a inhomogeneous liquid this formula is not exact. Corrections to this formula due
to forces from density gradients were found in Refs. \onlinecite{Ang} and \onlinecite{SR}. But these
corrections are essential only at the distance of the order $b$ from the sample boundary \cite{Cod} and
therefore are not essential for determination of eigenfrequencies until $R \gg b$.

\begin{figure}
  \begin{center}
    \leavevmode
    \includegraphics[width=\linewidth]{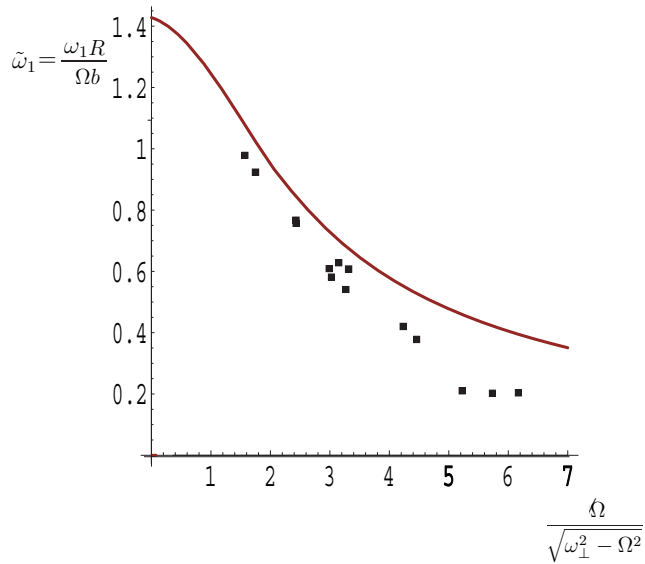}
    \caption{Comparison between the theory (solid line) and the experiment (black squares). }
  \label{fig1}
  \end{center}
  \end{figure}

In order to find the eigenfrequencies we determined numerically (at fixed $s$) eigenvalues of $q$ 
at which Eqs. (\ref{eqD1}) and  (\ref{eqD2}) have a solution with proper boundary conditions. In addition to
the boundary conditions at the BEC boundary discussed above, there are standard conditions in the BEC center that
the velocities are not divergent, namely, $v_r \propto v_t \propto r$. If compressibility is important any
reduced eigenfrequency is not a number but a function of $s$: $\tilde \omega_i = f_i(s)$. At $s \rightarrow 0$
the functions $f_i(s)$ are constant, but at large $s$ the functions $f_i(s)$ are inversely proportional to $s$.
The numerical solution  yields that  at large $s$  the two first eigenfrequencies  are
$\tilde \omega_1=7.17/s$  and  $\tilde \omega_2=16.9/s$. Figure \ref{fig1} shows the numerically found first
eigenfrequency $\tilde
\omega_1$  plotted as a function of
$\Omega/\sqrt{\omega_\perp^2-\Omega^2}=s/2\sqrt{2}$ (solid line).  The black squares show experimental points
\cite{exp} plotted in dimensionless variables by I. Coddington. They were obtained for various parameters, but
collapse on the same curve, as expected from the present analysis. Quantitative agreement between
the theory and the experiment looks quite good. It becomes worse at larger
$\Omega/\sqrt{\omega_\perp^2-\Omega^2}$, since the growth of the parameter $\xi/b$  in this area makes the
theory less accurate (see above). Coddington {\em al.} \cite{exp} measured also the  ratio of the two first
frequencies
$\omega_2/\omega_1 =1.8$  at $\Omega/\omega_\perp=0.95$, which corresponds to $s=8.61$, The present theory predicts
the ratio $\omega_2/\omega_1 =2.09$.

In summary the continuum theory of Tkachenko modes was developed, which takes into account the liquid
inhomogeneity and finite compressibility. Application of the theory to a rapidly rotating Bose--Einstein
condensate required formulation of proper boundary conditions at the condensate border where atom density
vanishes. The theory is in good agreement with observations of Tkachenko modes in the rapidly rotating
Bose--Einstein condensate.

I thank Ian Coddington who prepared the experimental part of Fig. \ref{fig1}. I also appreciate very much an
interesting comment  by Marco Cozzini, Lev Pitaevskii, and
Sandro Stringari  concerning comparison of the present analysis with that based on the compressibility sum
rule~\cite{Pit}.

\end{document}